\documentclass[12pt]{iopart}

\usepackage{iopams}
\usepackage{graphicx}
\usepackage{setspace}
\usepackage{mathrsfs}
\usepackage{subfig}
\usepackage[table]{xcolor}
\usepackage{hyperref}
\usepackage{enumitem}
\hypersetup{colorlinks, citecolor=black, filecolor=black, linkcolor=black, urlcolor=black}

\expandafter\let\csname equation*\endcsname\relax
\expandafter\let\csname endequation*\endcsname\relax
\usepackage{amsmath}

\definecolor{grayish}{RGB}{230,230,230}

\newcommand{\refEq}[1] {(\ref{#1})}

\newcommand{\romanNum}[1]{\uppercase\expandafter{\romannumeral#1}}

\newcommand{\elem}[2] {\ensuremath{^{#2}\text{#1}}}

\begin{document}

\title{Maximizing specific energy by breeding deuterium}

\author{Justin Ball}

\address{Ecole Polytechnique F\'{e}d\'{e}rale de Lausanne (EPFL), Swiss Plasma Center (SPC), CH-1015 Lausanne, Switzerland}

\ead{Justin.Ball@epfl.ch}

\vspace{2pc}

\begin{abstract}
	
	Specific energy (i.e. energy per unit mass) is one of the most fundamental and consequential properties of a fuel source. In this work, a systematic study of measured fusion cross-sections is performed to determine which reactions are potentially feasible and identify the fuel cycle that maximizes specific energy. This reveals that, by using normal hydrogen to breed deuterium via neutron capture, the conventional catalyzed D-D fusion fuel cycle can attain a specific energy greater than anything else. Simply surrounding a catalyzed D-D reactor with water enables deuterium fuel, the dominant stockpile of energy on Earth, to produce as much as 65\% more energy. Lastly, the impact on space propulsion is considered, revealing that an effective exhaust velocity exceeding that of deuterium-helium-3 is theoretically possible.
	
\end{abstract}

\section{Introduction}
\label{sec:intro}

The aim of this work is to investigate the basic question ``What fuel source can provide the most energy with the least mass?'' For this question, we will look into the near future and assume that the current incremental and steady development of technology will continue. However, we will not postulate any breakthroughs or new physical processes because they cannot be predicted with any certainty. It is already well-known that the answer to our question must involve fusion fuels \cite{EnglertFirstDHe3spacecraft1962, ZubrinEnteringSpace1999}, but nothing more precise has been rigorously established\footnote{Matter-antimatter annihilation does have a higher specific energy than fusion, but it appears impractical for two reasons. First, antimatter does not occur naturally in our Solar System \cite{SteckerAntimatter1971} and is inefficient to produce. It currently requires over 10 million units of input energy to create enough antimatter to generate 1 unit of energy. Second, it is difficult to store because it annihilates with normal matter. The most that has been confined by a device corresponds to just 300 Joules \cite{ZubrinEnteringSpace1999}.}. A more precise answer may be useful in the future when the technology required to implement it becomes available. Specific energy (i.e. energy per unit mass) is such a basic physical property that it seems almost certain to have important applications. However, perhaps more importantly, the answer to such a fundamental question is also of intrinsic value, simply to satisfy our curiosity --- to know.

In this work, we will first introduce the conventional fusion fuel cycles, focusing on the catalyzed D-D cycle. Then, in sections \ref{sec:breedingDeut} and \ref{sec:implementation} we will propose a simple extension to it using the neutron capture reaction by normal hydrogen and give an example of how it could be implemented. Next, in section \ref{sec:specEnergyProof} we will argue that this fuel cycle possesses the maximum specific energy of any cycle that appears even remotely possible. Lastly, section \ref{sec:propulsion} will consider the fuel cycle's utility for spacecraft propulsion and section \ref{sec:conclusions} will provide some concluding remarks.

\section{Fusion fuel cycles}
\label{sec:fusionIntro}

When fusion energy research started in the 1950s, a great deal of attention was given to fusion fuel cycles in order to determine which was optimal for use in future power plants \cite{PostLectures1954, SimonSherwoodIntro1955, PostFusionIntro1956, MileyAltFuelsOverview1980}. After a short period of hesitation \cite{SimonSherwoodIntro1955, BelloFortune1957,  RobinsonBreedingIssues1958}, the deuterium-tritium (D-T) fusion reaction,
\begin{align}
  \elem{H}{2} + \elem{H}{3} \rightarrow \elem{He}{4} + \elem{n}{1} + \text{17.6 MeV} ,  \label{eq:DTreaction}
\end{align}
was identified as the most promising reaction \cite{SpitzerStellarators1954, DavisTritiumBreeding1958}. It is optimal because it has the largest fusion cross-section ($\sim$5 barns) and this occurs at a relatively low reactant energy ($\sim$60 keV). However, since tritium has a half-life of just 12.3 years, it is not abundant in nature. Because of this, it is generally envisioned to be produced within the power plant, rather than being an input. This can be done by surrounding the reactor with a ``blanket'' that includes lithium, in order to capture the neutron produced by D-T fusion to ``breed'' tritium, e.g.
\begin{align}
  \elem{Li}{6} + \elem{n}{1} \rightarrow \elem{H}{3} + \elem{He}{4} + \text{4.8 MeV} .  \label{eq:Tbreeding}
\end{align}
Given a small amount of tritium for start-up, this ``breeding'' reaction closes the overall D-T fuel cycle,
\begin{align}
  \elem{H}{2} + \elem{Li}{6} \rightarrow 2 ~ \elem{He}{4} + \text{22.4 MeV} , \label{eq:DTfuelCycle}
\end{align}
because both deuterium and lithium-6 are abundant in nature.

The deuterium-helium-3 reaction (D-He-3) is generally considered to be the next most promising reaction. It has a maximum cross-section of $\sim$0.8 barns at $\sim$200 keV and is given by
\begin{align}
  \elem{H}{2} + \elem{He}{3} \rightarrow \elem{He}{4} + \elem{H}{1} + \text{18.4 MeV} .  \label{eq:DHe3reaction}
\end{align}
It is attractive, but helium-3 is fairly rare on Earth \cite{WittenbergHe3Resources1993} and the reaction does not produce a neutron with which to breed it. Thus, this reaction requires helium-3 to be found elsewhere (e.g. the moon \cite{WittenbergHe3Resources1993, WittenbergDHe3Review1992}) or to be produced using other nuclear reactions \cite{WittenbergHe3Resources1993, MileyDHe3sources1988}.

The next best fusion reaction occurs between two deuterium nuclei (D-D). While the maximum cross-section is just $\sim$0.2 barns, which occurs at $\sim$1000 keV, this peak is very broad. This means that significant fusion power can be generated at temperatures much lower than 1000 keV. Interestingly, the D-D reaction has two branches that occur with almost equal probability,
\begin{align}
  2 ~ \elem{H}{2}
  \begin{array}{l}
    \rotatebox[origin=r]{20}{$\to$} \elem{H}{3} + \elem{H}{1} + \text{4.0 MeV} \\
    \rotatebox[origin=r]{-20}{$\to$} \elem{He}{3} + \elem{n}{1} + \text{3.3 MeV.}
  \end{array} \label{eq:DDreaction}
\end{align}
Note that, while the previous two reactions produce the exceptionally stable helium-4 nucleus, this reaction produces tritium and helium-3. This allows for the possibility of the ``catalyzed'' D-D fuel cycle \cite{SimonSherwoodIntro1955, MillsCatDD1971, MileyDDtokamaks1977}, where the tritium and helium-3 produced by D-D fusion are fused with deuterium in subsequent reactions. This can be done in the same device or the products can be extracted and fused in a separate devices. Regardless, the overall catalyzed D-D fuel cycle is given by the sum of equations \refEq{eq:DTreaction}, \refEq{eq:DHe3reaction}, and \refEq{eq:DDreaction}:
\begin{align}
  3 ~ \elem{H}{2} \to \elem{He}{4} + \elem{H}{1} + \elem{n}{1} + \text{21.6 MeV} . \label{eq:catalyzedDDreaction}
\end{align}
Such a fuel cycle performs significantly better than D-D reactions on their own. In fact, the temperature required to achieve a self-sustaining fusion burn with catalyzed D-D is lower than for D-He-3 and the fusion power density of the plasma is higher (though the performance is still significantly worse than D-T) \cite{MillsCatDD1971, MileyDDtokamaks1977}.

These are the only reactions that currently appear to have practical application \cite{RiderAltFuels1995, HiltonFusionPropulsion1964}. All other fusion reactions are more difficult. They have lower fusion cross-sections that occur at higher temperatures and/or involve reactants with higher electric charge. Because of this, the power lost by electron bremsstrahlung looks to exceed the fusion power for nearly all other fusion reactions \cite{RiderAltFuels1995}. This is a fairly fundamental constraint because bremsstrahlung radiation is emitted whenever charged particles collide, which is intrinsic to the process of fusion. Moreover, bremsstrahlung is generally in the ultraviolet/X-ray range of frequencies, so it looks difficult to reflect, reabsorb in the plasma, or convert to electricity at high efficiency \cite{RiderAltFuels1995}. Unless this problem can be addressed (e.g. maintaining cold electrons \cite{LernerQuantumMagFieldEffect2008}, removing electrons \cite{HuangNonNeutral1997}, non-equilibrium plasmas \cite{RostokerCBFR1997, HoraNonThermalPB2018}), it does not appear possible to produce net electricity using nearly all other fusion reactions. At the present time, even the D-T fusion reaction proves very challenging, with current fusion devices still somewhat short of net energy production \cite{KeilhackerJETrecord1999}.

\section{Breeding deuterium}
\label{sec:breedingDeut}

This work proposes a simple extension to the catalyzed D-D fuel cycle: to use normal hydrogen (i.e. protium) to breed deuterium with the neutron capture reaction,
\begin{align}
  \elem{H}{1} + \elem{n}{1} \to \elem{H}{2} + \text{2.2 MeV} . \label{eq:Dbreeding}
\end{align}
This reaction has a cross-section of $\sim0.3$ barns for thermal neutrons \cite{HamermeshDbreedingCrossSection1953}. Though the cross-section may seem small compared to D-T fusion, this reaction is actually much easier to accomplish. Because it occurs at room temperature and neutrons don't interact much with electrons, you can place a liquid (or even a solid) containing normal hydrogen in the path of neutrons. Liquids have a much greater number density than the plasmas typical of fusion, so the fact that the individual nuclei have somewhat smaller cross-sections can be overcome. The neutrons will enter the material, thermalize due to scattering collisions, and be captured.

Using the neutron capture reaction enables the fuel cycle
\begin{align}
  2 ~ \elem{H}{2} \to \elem{He}{4} + \text{23.8 MeV} . \label{eq:catalyzedDDwithBreeding}
\end{align}
This set of reactions will be referred to as the ``catalyzed D-D+D'' fuel cycle. While this idea seems fairly obvious from equation \refEq{eq:catalyzedDDreaction}, the practicality and advantage of deuterium breeding does not appear to have been noted. For example, references \cite{EnglertFirstDHe3spacecraft1962, ZubrinEnteringSpace1999, SantariusDHe3Spacecraft1992} all conclude that D-He-3 has the highest specific energy of any fuel. However, the specific energy of the catalyzed D-D+D cycle is more than 60\% larger (see table \ref{tab:specEnergy})\footnote{Since D-He-3 releases all of its energy in the form of charged particles that have the potential to be converted to electricity at a high efficiency, one might think that D-He-3 could have a higher specific energy in practice. However, catalyzed D-D+D releases 3.4 MeV/AMU in charged particles alone, already almost equal to that of D-He-3. If the energy carried by neutrons can be made useful at even a low efficiency, catalyzed D-D+D will be more energy dense.}. Moreover, the plasma performance would be identical to the normal catalyzed D-D (which is seen as one of the most viable fusion fuel cycle after D-T) because the only changes we have introduced occur outside the plasma. Additionally, capturing the fusion neutrons with normal hydrogen prevents them from activating other surrounding material, thus minimizing the production of radioactive waste.

\begin{table}
	\centering
	\caption{The specific energy of some of the easiest fusion fuel cycles.}
	\begin{tabular}{ l c}
		Fuel cycle & Specific energy \\
		\hline
		Catalyzed D-D+D & 6.0 MeV/AMU \\
		D-He-3 & 3.7 MeV/AMU \\
		Catalyzed D-D & 3.6 MeV/AMU \\
		D-T & 3.5 MeV/AMU \\
		D-T w/ T breeding & 2.8 MeV/AMU
	\end{tabular}
	\hspace{3em}
	\begin{tabular}{ l c}
		Fuel cycle & Specific energy \\
		\hline
		D-Li-6 & 2.8 MeV/AMU \\
		He-3-He-3 & 2.2 MeV/AMU \\
		T-T & 1.8 MeV/AMU \\
		D-D & 0.9 MeV/AMU \\
		p-B-11 & 0.7 MeV/AMU
	\end{tabular}
	\label{tab:specEnergy}
\end{table}

Assuming the ability to achieve a conventional catalyzed D-D fuel cycle, the addition of deuterium breeding looks to be attractive and easily achievable. It can increase the specific energy of deuterium fuel by as much as 65\% relative to a catalyzed D-D fuel cycle {\it without} deuterium breeding. Put another way, any given deuterium fuel supply could be used to generate as much as 65\% more energy, without any additional inputs. This could be useful for a number of future applications. First, deuterium is the third most abundant isotope in the universe \cite{OmearaDeuteriumAbundance2001} and one in every 6420 hydrogen atoms on Earth is deuterium. Moreover, the two most abundant isotopes, hydrogen-1 and helium-4, appear impossible to fuse without using other less abundant isotopes. Helium-3 can be fused and is almost as abundant as deuterium in the universe, but not in our neighborhood. It is nearly non-existent on Earth and, even when totaled over all the planets in our Solar System, contains just 1/50 of the energy stored by deuterium in the Earth's oceans \cite{ZubrinEnteringSpace1999}. Thus, deuterium appears to be our dominant usable stockpile of energy. By partially breeding it, our supply can be used most efficiently.

Second, an ultra-high specific energy fuel can be advantageous in any situation that requires fuel to be transported. A clear example is industrial activities at stationary, remote outposts in outer space. For example, asteroid mining, transmission stations, or large construction projects could be best fueled by catalyzed D-D+D. Such activities would require a lot of energy, but be insensitive to the mass and volume of the power plant (especially if the raw material for the plant could be found on-site). Another particularly important application could be space propulsion, which we will consider in a later section. Of course, all these applications are sufficiently far in the future that we can only speculate. Regardless, specific energy is such a fundamental property that the fuel that maximizes it seems likely to have important applications.

\section{An illustrative example}
\label{sec:implementation}

\begin{figure}
	\centering
	\includegraphics[width=0.6\textwidth]{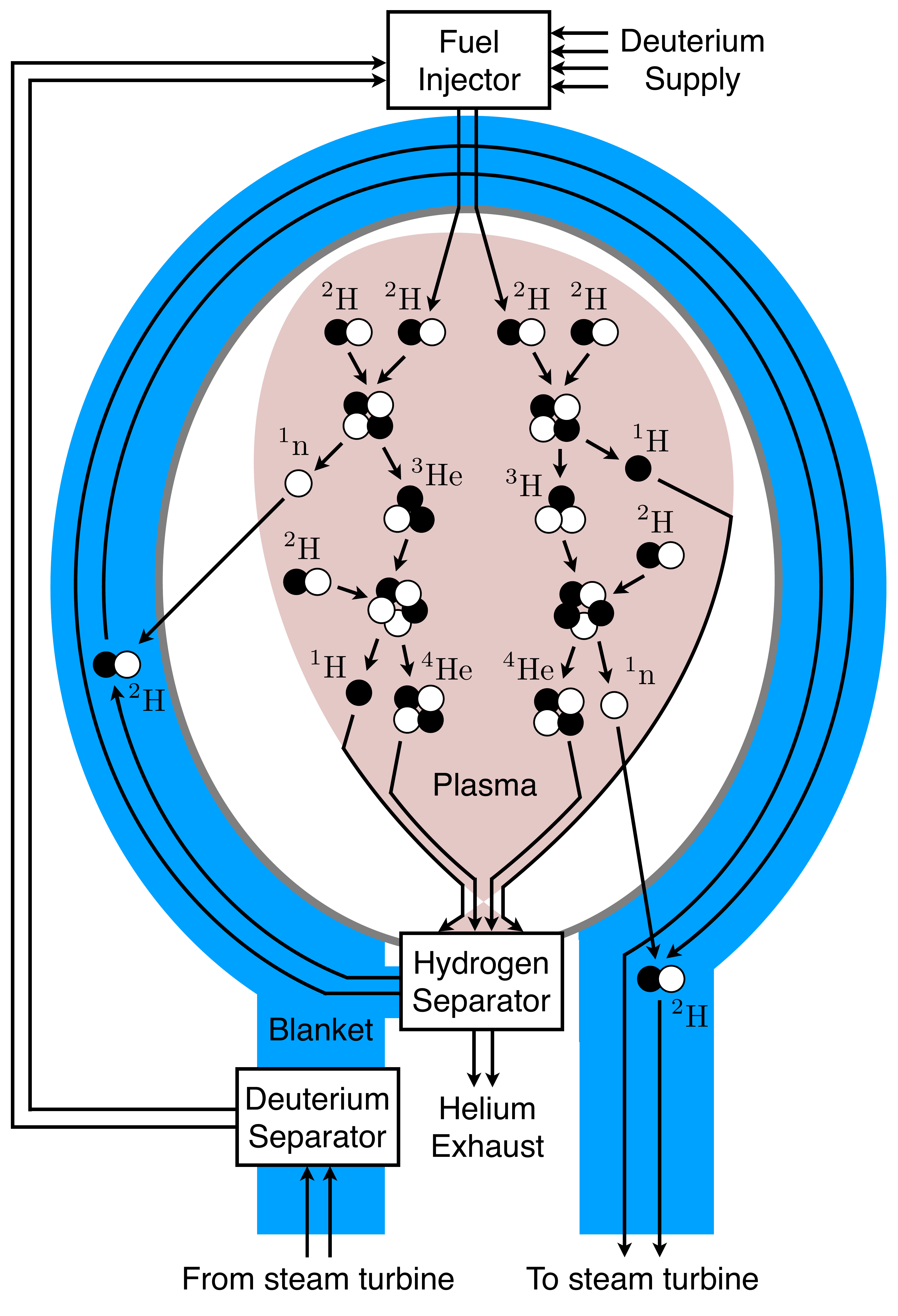}
	\caption{A cartoon of a power plant that implements the catalyzed D-D+D fuel cycle.}
	\label{fig:reactorSchematic}
\end{figure}

\begin{figure}
  \centering
  \includegraphics[width=0.7\textwidth]{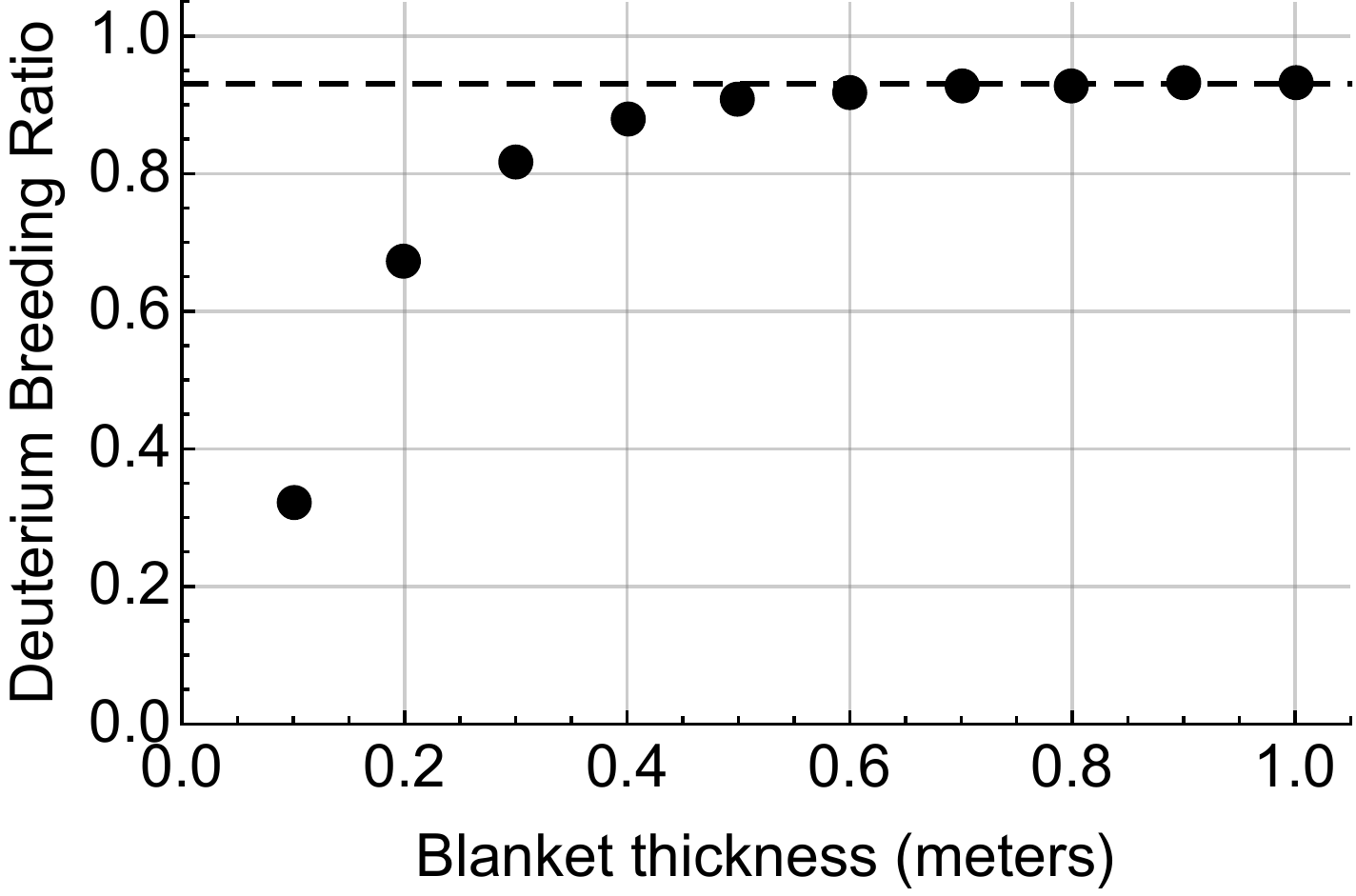}
  \caption{The deuterium breeding ratio as a function of the blanket thickness, where the horizontal dashed line gives the value in the limit of a very thick blanket.}
  \label{fig:DBR}
\end{figure}

A simple method to implement the catalyzed D-D+D fuel cycle is to surround the D-D and D-T reactions with a blanket of normal light water. This could be pressurized, similar to a pressurized water fission reactor (PWR), or allowed to boil, similar to a boiling water fission reactor (BWR). Figure \ref{fig:reactorSchematic} illustrates such a configuration using an ITER-like tokamak as an example confinement scheme \cite{ShimadaITER2007}, but it could just as well be an inertial confinement scheme or something more exotic. The only constraint is that there be space for the deuterium breeding blanket, which figure \ref{fig:DBR} shows would have a similar thickness to the {\it tritium} breeding blankets in current D-T power plant designs \cite{RaffrayARIESpowerCore2006, SorbomARC2015}. Using light water makes for an attractive design because it is the most common coolant in power plants and has already been used extensively in the neutron environment of light water fission reactors. Hence, its neutronic properties are well studied. Nuclear fission reactors have been observed to generate kilograms of deuterium per year. The only significant difference for a fusion reactor is the energy of the neutrons --- fission neutrons are born with $\sim 1$ MeV of energy while D-T neutrons have 14.1 MeV and D-D neutrons have 2.5 MeV. Nevertheless, the cross-sections of the neutron-induced reactions of hydrogen-1 and oxygen-16 are well characterized at these energies.

To demonstrate the feasibility of such a design, neutronics simulations were performed using MCNP \cite{MCNP2003}. An axial source of D-D and D-T neutrons (weighted equally) was entirely surrounded by thick cylindrical blanket of liquid light water at atmospheric pressure and temperature. The deuterium breeding ratio (i.e. the number of deuterium nuclei produced per fusion neutron) is shown in figure \ref{fig:DBR} for different blanket thicknesses. Note that oxygen-16 reacts with a very small fraction of the neutrons because it a double magic isotope, so it is exceptionally stable. Lastly, in reality a design would likely have solid metal plasma-facing components in front of the blanket. This structure would absorb some of the neutrons before they can reach the hydrogen. While this isn't ideal, unlike tritium breeding for the D-T fuel cycle, it is perfectly acceptable for the DBR to be less than one. If some of the neutrons are lost then the fuel cycle will require more deuterium as input, but the overall fuel cycle will still function. Note that using material to multiply neutrons (e.g. beryllium or lead) could be used, but these are substantially less abundant than deuterium and would only decrease the specific energy of the fuel cycle as a whole because of their large atomic mass.

In the design shown in figure \ref{fig:reactorSchematic}, the liquid light water serves as both a coolant and a breeding blanket. The fusion reaction chamber is submerged in water, which slowly circulates past. This is similar to breeding blankets that have already been proposed for D-T power plants \cite{SorbomARC2015}. As it passes by the tank, the water intercepts the neutrons, causing it to heat up and produce deuterium. The heated water can then be used to power a steam cycle and generate electricity. Afterwards, the deuterium could be separated from the normal water by the Girdler Sulfide process, which is currently used to extract deuterium from seawater. Periodically, normal hydrogen collected from the plasma could be combined with the oxygen leftover from the deuterium extraction. This would be added to the coolant. Unlike in the D-T fuel cycle, the speed of such a deuterium breeding cycle is not particularly important. In D-T power plants, the time required to breed, extract, and fuse the tritium determines the total power plant tritium inventory, which must be minimized for safety reasons. Deuterium, on the other hand, can be allow to accumulate for a long time and does not disappear due to radioactive decay.

Finally, it should be mentioned that, for any particular application, there are many competing design considerations and the specific energy of the fuel is just one. That is to say that the high specific energy of the catalyzed D-D+D fuel cycle is only useful to the extent that the specific energy of the {\it fuel} actually matters. For many systems, specific energy is only a secondary consideration compared to factors like neutron activation, plasma power density, and the required fusion technology. Indeed for many applications, any fusion fuel already has such high specific energy that it can be negligible compared to other necessary components. For example, in present-day fusion power plant designs \cite{SorbomARC2015}, a water blanket as depicted above could easily have a mass 1000 times greater than the mass of deuterium that it would breed each year. Conceptual designs of D-He-3 spacecraft thrusters for travel within the Solar System \cite{HiltonFusionPropulsion1964, SantariusDHe3Spacecraft1992}, which are designed to minimize mass, are still over 100 times more massive than the fuel they use. Of course, fuel mass would be more significant in devices that are larger, have higher power densities, and longer lifespans. Such devices would hopefully be developed in the future and would likely be required for interstellar space travel. Moreover, there may be situations in which normal hydrogen and/or structural materials are abundant, but fusion fuel is difficult to access.

\section{Maximizing specific energy}
\label{sec:specEnergyProof}

To argue that the catalyzed D-D+D fuel cycle (i.e. equation \refEq{eq:catalyzedDDwithBreeding}) maximizes specific energy, we first note that nuclear forces are more than a million times stronger than the other two fundamental forces. The gravitational potential energy of two nuclei, even if they are just a femtometer apart, is less than $10^{-25}$ eV. Electromagnetism is responsible for chemical bonds between atoms, the strongest of which are roughly 10 eV. Nuclear reactions, on the other hand, typically release millions of eV. Hence, the reactions with the highest specific energy seem almost certain to be nuclear.

\begin{figure}
  \centering
  \includegraphics[width=\textwidth]{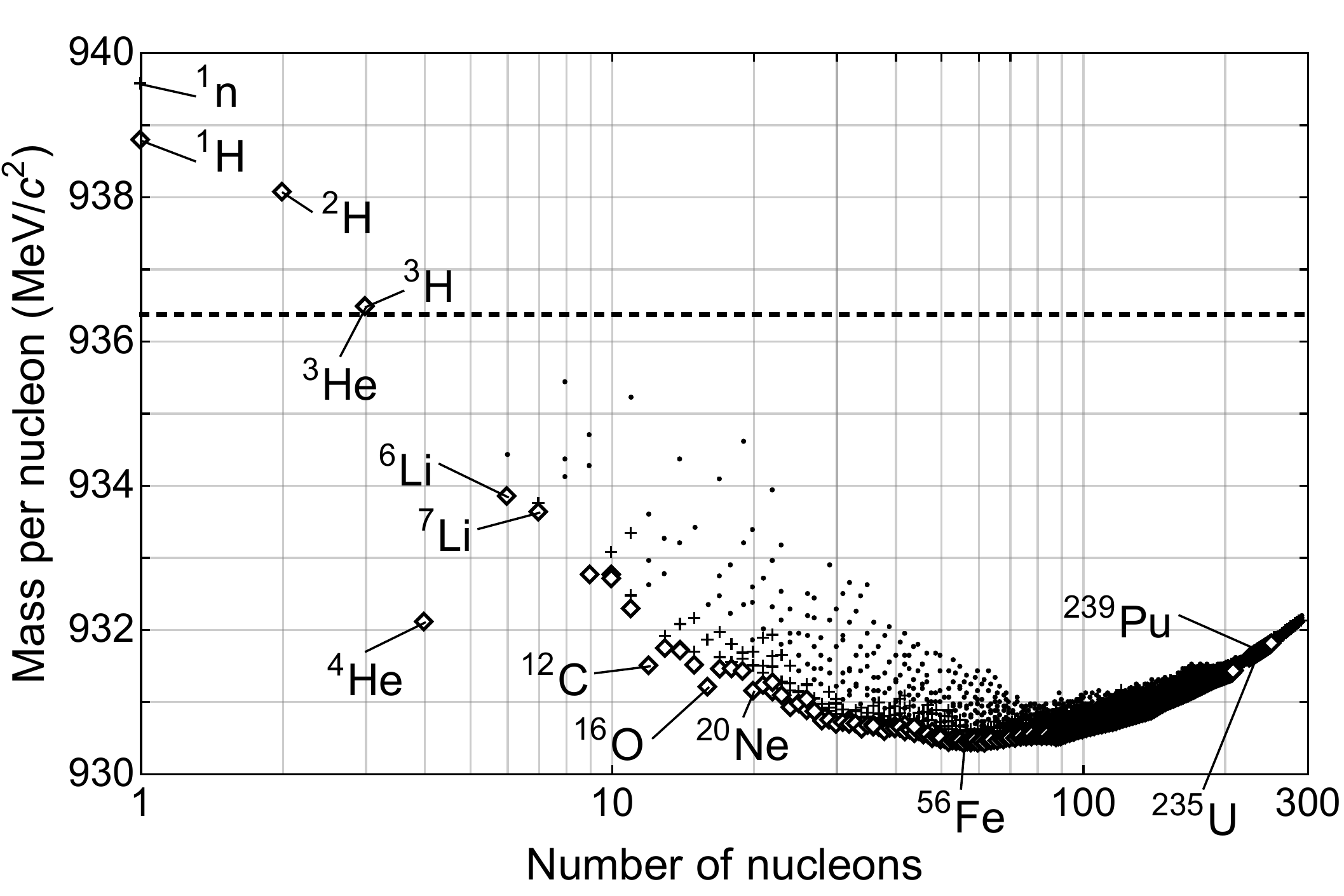}
  \caption{The rest mass per nucleon for all isotopes with half-lives longer than a year (diamonds), a second (crosses), and a microsecond (dots). The value of iron-56 plus the specific energy of the catalyzed D-D+D cycle is indicated by the horizontal dashed line.}
  \label{fig:BEcurve}
\end{figure}

The energetics of nuclear reactions are well summarized by the mass per nucleon curve shown in figure \ref{fig:BEcurve}. Due to $E = m c^{2}$, converting reactants with high mass nucleons to products with low mass nucleons will release energy. More precisely, {\it the specific energy of a nuclear reaction is well approximated by the average mass of a nucleon in the reactants minus the average mass of a nucleon in the products} (times the speed of light squared). Strictly speaking, this actually gives the energy released {\it per nucleon}, rather than the specific energy (i.e. the energy released {\it per mass}). However, figure \ref{fig:BEcurve} shows that these two quantities are identical to within 1\%. Regardless, we would like to use reactants near the top of this plot to make products near the bottom.

From figure \ref{fig:BEcurve}, we see that converting two deuterons into one helium-4 releases almost 6 MeV/AMU, which spans over 70\% of what is theoretically attainable (i.e. the reaction $56 ~ \elem{H}{1} \rightarrow \elem{Fe}{56} + \text{8.4 MeV/AMU}$). Intuitively, this seems hard to beat, especially considering how difficult fusion reactions are. Even if the only product of a fuel cycle is the most energetically favorable isotope (i.e. iron-56), the reactants still must have an average mass per nucleons greater than 936.4 MeV/AMU to beat catalyzed D-D+D. By this logic, we can rule out all fission processes. The input nucleons are too light (see figure \ref{fig:BEcurve}).

Thus, we can restrict our considerations to fusion fuel cycles (supplemented by other nuclear reactions like neutron capture and radioactive decay). At this point, things become a bit more subjective because we must decide which fusion reactions are practical to achieve. Converting 56 protons into iron-56 has the highest specific energy, but it has only been accomplished over millions of years using a star and looks absolutely impossible to do on any sort of human scale. However, there could be other fairly viable fusion reactions that do better than catalyzed D-D+D. This is a messy empirical question, so we will perform a systematic study of ``the most comprehensive compilation of experimental nuclear reaction data'' --- the EXFOR database \cite{OtukaEXFOR2014}.

To estimate the difficulty of various fusion reactions we will use a variant of a common fusion metric: the triple product required for ignition \cite{WessonTokamaks2004}. The triple product $n T_{i} \tau_{E}$ is the product of the electron density, ion temperature, and energy confinement time. In order for a fuel to ignite, it must achieve a minimum triple product given by \cite{LawsonCriterion1957}:
\begin{align}
   n T_{i} \tau_{E} = 3 \left( 2 Z_{1} Z_{2} + Z_{1} + Z_{2} \right) \frac{T_{i}^{2}}{\left\langle \sigma v_{i} \right\rangle} \frac{1}{f_{c} E_{fus}} . \label{eq:tripleProd}
\end{align}
The lower this value is, the easier the fusion reaction is. Here $Z_{1}$ and $Z_{2}$ are the charge numbers of the two reactants, $\sigma$ is the fusion reaction cross-section, $v_{i}$ is the relative speed between the two reactants, $\left\langle \ldots \right\rangle$ is an average over the ion velocity distributions, $E_{fus}$ is the energy released by the fusion reaction, and $f_{c}$ is the fraction of it that is carried by charged particles. To calculate this we have assumed that both of the ion densities are $n_{i} = n / (2 Z_{i} )$ because this choice maximizes the fusion power density. Note that if the two reactants are identical (e.g. D-D fusion), the prefactor becomes $3 Z_{1} \left( Z_{1} + 1 \right)$ and the ion density is required to be $n_{1} = n / Z_{1}$ by the quasineutrality condition.

For our purposes, we will make a few adjustments to equation \refEq{eq:tripleProd}. First, we will multiply by $f_{c}$ because we are primarily concerned about the total energy released, rather than details about what energy is deposited in the plasma and how efficiently power can be recirculated. Second, because the EXFOR database contains the results of many independent experimental measurements, the cross-section values are not evenly distributed in energy. Many unusual reactions have just a few points. Fortunately, we only require a rough estimate of the difficulty of a given reaction. Thus, we will calculate the ``point'' triple product by assuming that the ion distribution function is a delta function at a given center-of-mass energy $E_{i}$. With these assumptions, equation \refEq{eq:tripleProd} becomes
\begin{align}
n E_{i} \tau_{E} f_{c} = 3 \left( 2 Z_{1} Z_{2} + Z_{1} + Z_{2} \right) \frac{E_{i}^{2}}{\sigma \left( E_{i} \right) \sqrt{2 E_{i}/m_{i}}} \frac{1}{E_{fus}} , \label{eq:pointTripleProd}
\end{align}
where $m_{i}$ is reduced ion mass of the reactants. This metric applies equally well to a dataset with one point or a thousand. While this is a simple and useful metric to give a rough estimate, it is not perfect. It will tend to overrate fusion reactions that have sharp, high resonances in their cross-sections and underrate reactions with broad, low cross-sections (e.g. D-D). Additionally, it doesn't make much sense for endothermic reactions because a larger magnitude of $E_{fus}$ is a bad thing. Thus, we will compute $n E_{i} \tau_{E} f_{c} E_{fus}^{2}$ for endothermic reactions.

\begin{figure}
	\centering
	\includegraphics[width=\textwidth]{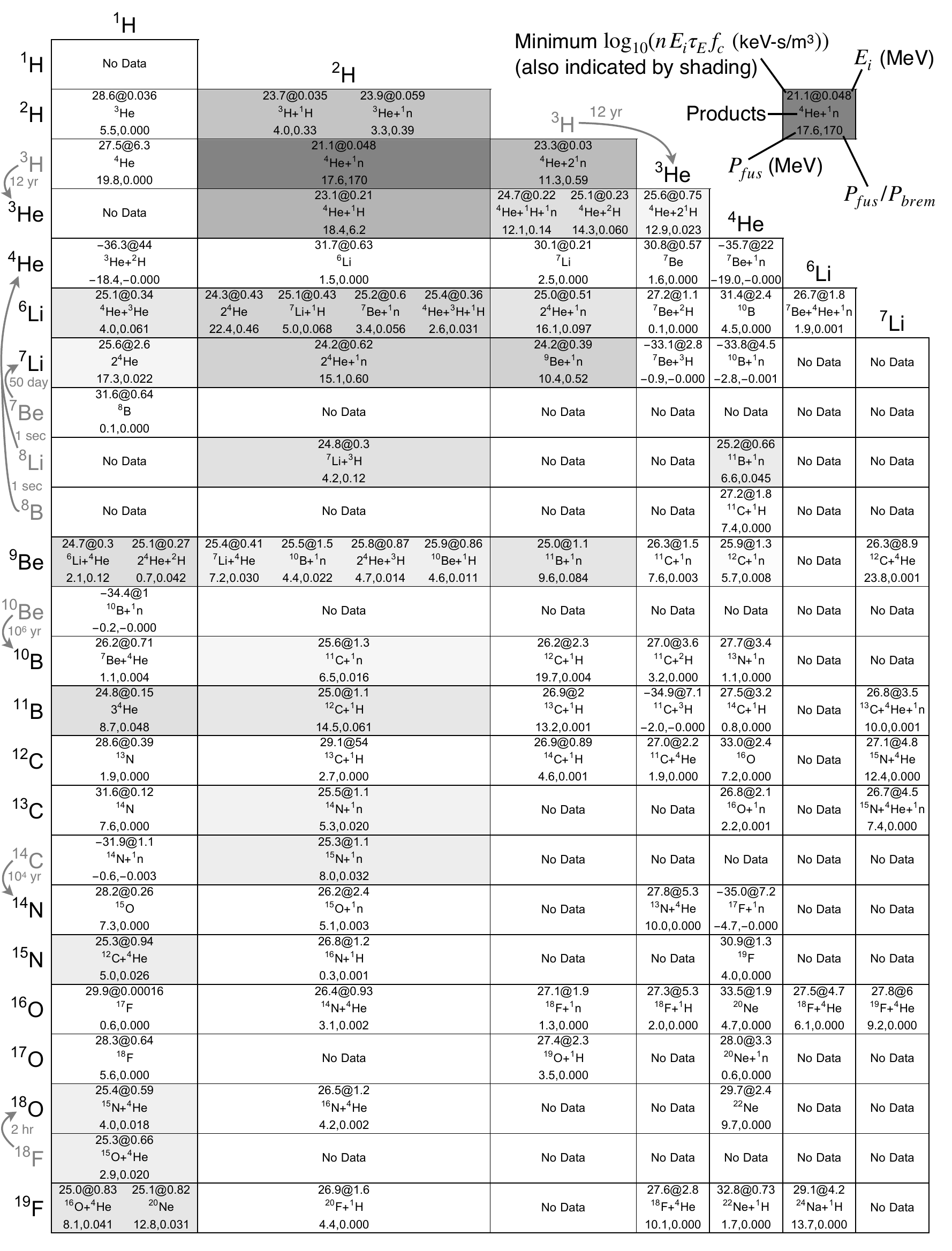}
	\caption{A summary of all measured fusion reactions in the EXFOR database \cite{OtukaEXFOR2014} between light nuclei. The stable (black) and unstable (gray) reactants are given by the column and row labels, while the legend in the upper-right corner explains the contents of each reaction cell. Negative values indicate endothermic reactions (see text).}
	\label{fig:possibleReactions}
\end{figure}

Figure \ref{fig:possibleReactions} shows these values for the complete contents of the EXFOR database. The column and row labels indicate the two reactants of the fusion reaction. The table boundaries were chosen such that it includes all fusion reaction with $n E_{i} \tau_{E} f_{c} < 10^{26}$ keV-s/m\textsuperscript{3}. This number was chosen as the cutoff for potentially feasible reactions because it is more than a factor of 10 above p-B-11, the hardest reaction that some present-day researchers are pursuing \cite{LernerQuantumMagFieldEffect2008, RostokerCBFR1997, HoraNonThermalPB2018}. Isotopes with no data were omitted. The first number of the first row within each cell (as well as the shading of the cell) indicates the minimum value of the point triple product across all energy values and datasets in the database. Specifically, the number is $\log_{10} \left( n E_{i} \tau_{E} f_{c} \right)$, where $n E_{i} \tau_{E} f_{c}$ is expressed in units of keV-s/m\textsuperscript{3}. The number that follows the ``@'' sign is the ion energy (in MeV) that this minimum value occurs at. Easier fusion reactions are indicated by both of these numbers being smaller. The second line gives the products that are generated by the reaction. Any product that has a half-life of less than 1 second has been replaced by its daughter products, unless it appears as a reactant on the chart. The first number in the third line indicates the fusion energy (in MeV) released by the reaction. The number after the comma is the ratio of the fusion power density to the bremsstrahlung power density radiated by electrons. This is calculated with the formulas \cite{RiderAltFuels1995, NRLformulary}
\begin{align}
   p_{\text{fusion}} &= n_{1} n_{2} \sigma \left( E_{i} \right) \sqrt{2 E_{i}/m_{i}} E_{fus} , \label{eq:fusPowerDensity}	\\
   p_{\text{brem}} &= \frac{160}{3} \left( \frac{e^{2}}{4 \pi \epsilon_{0}} \right)^{3} \frac{n^{2} \sqrt{E_{i}}}{\left( m_{e} c^{2} \right)^{3/2} \hbar} \\
   &\times \left( \frac{\sum_{i} Z_{i}^{2} n_{i}}{n} \left( 1 + 0.79 \frac{E_{i}}{m_{e} c^{2}} + 1.9 \left( \frac{E_{i}}{m_{e} c^{2}} \right)^{2} \right) + \frac{3}{\sqrt{2}} \frac{E_{i}}{m_{e} c^{2}} \right) .	\nonumber
\end{align}
Here $e$ is the electric charge of a proton, $\epsilon_{0}$ is the vacuum permittivity, $m_{e} c^{2}$ is the electron rest energy, and $\hbar$ is the reduced Planck constant. Note that for simplicity, the electron {\it temperature} has been assumed to be equal to the ion {\it energy}. The ion densities $n_{1}$ and $n_{2}$ are again determined in order to maximize the fusion power density, while respecting quasineutrality. If there is only one ion species present the factor of $n_{1} n_{2}$ in equation \refEq{eq:fusPowerDensity} becomes $n_{1}^{2}/2$ and the bremsstrahlung formula remains unchanged. Note that easier fusion reactions have larger numbers in the third line.

Lastly, some cells have more than one entry. This indicates that the reaction has multiple branches that are exothermic and have $n E_{i} \tau_{E} f_{c} < 10^{26}$ keV-s/m\textsuperscript{3}. If the reaction has several equally probable branches (e.g. D-D), then combining them increases the cross-section by a factor of $N_{branch}$, the number of branches. To account for this in the logarithm of the point triple product, you should subtract $\log_{10} ( N_{branch} )$. Additionally, it is appropriate to sum the different fusion-to-bremsstrahlung power ratios as the fusion power will increase due to the multiple reactions, but the power radiated by the electrons will not change.

The most eye-catching feature of figure \ref{fig:possibleReactions} is the number of reactions that have no data. In some cases, this is because the cross-section is too small to be experimentally measured (e.g. p-p, p-He-3 \cite{AdelbergerSolarCrossSections1998}) or one of the reactants has a short half-life (e.g. Li-8, B-8). However, all these unknowns should be kept in mind as they make the conclusions of this work somewhat less certain. Other causes for uncertainty include the threshold that has been set at $n E_{i} \tau_{E} f_{c} < 10^{26}$ keV-s/m\textsuperscript{3}. If we get better than this at fusion (or can efficiently accomplish endothermic reactions), there could be useful reactions that don't appear in the figure. Additionally, some of the point triple product numbers can be changed somewhat. This is because we have made certain assumptions about what unstable products we retain or let decay (as described above). These choices affect the energetics of the reaction and can alter the triple product. The effect of this has been checked by repeating the analysis while retaining all products with a half-life greater than 1 microsecond. The conclusions that will be presented were not affected.

\begin{figure}
	\centering
	\includegraphics[width=0.7\textwidth]{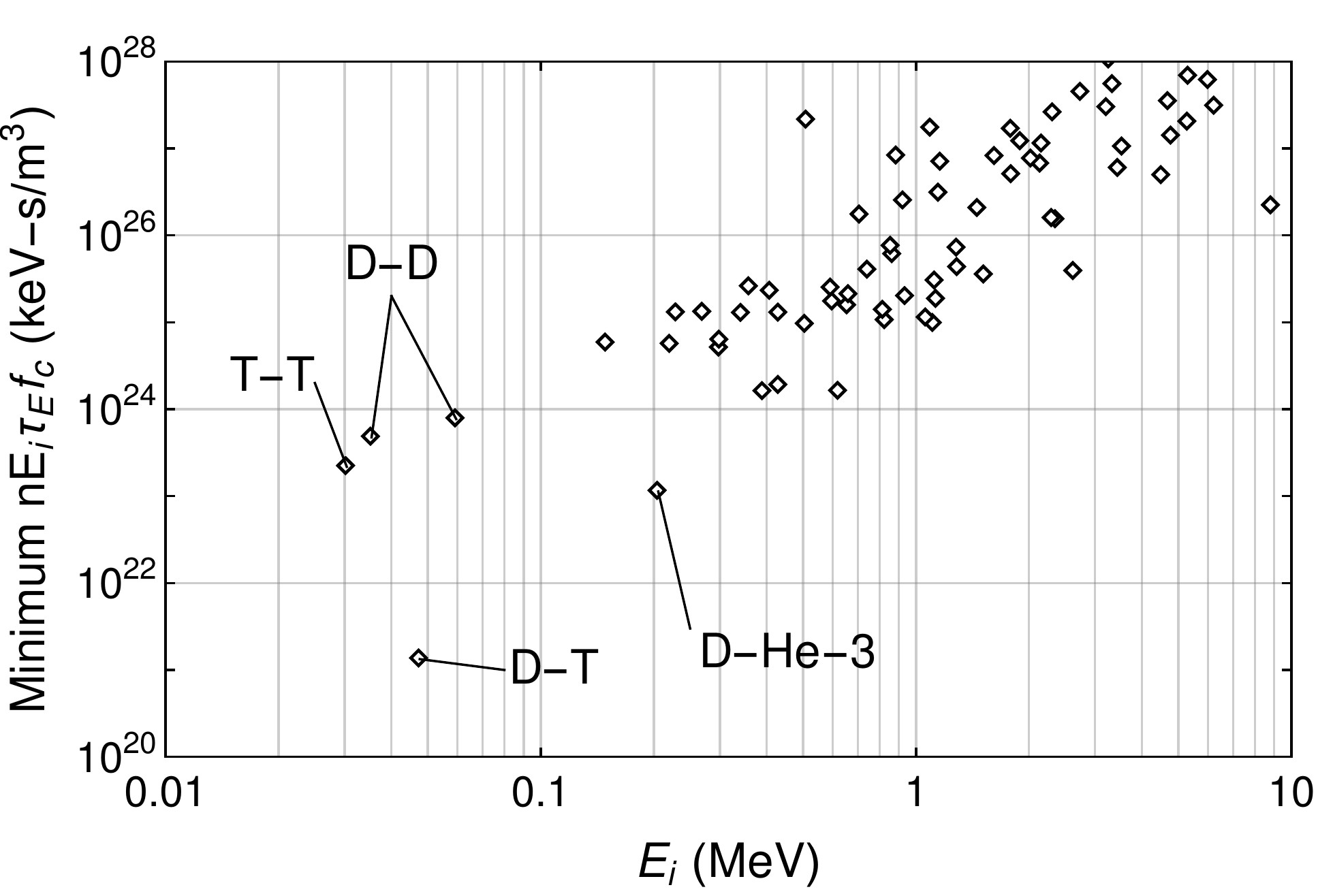}
	\caption{The minimum value of the point triple product for the data of figure \ref{fig:possibleReactions}.}
	\label{fig:minTripleProd}
\end{figure}

Digging deeper into figure \ref{fig:possibleReactions}, we see some expected results. D-T is the easiest reaction to achieve ignition by a substantial margin and D-He-3 is next. D-D appears to be harder than T-T, but when you include both branches of D-D in the same point triple product by subtracting $0.3$, you see that they are roughly even. Moreover, T-T fusion releases a much higher fraction of its energy in the form of neutrons, which is harder to utilize. All other reactions are significantly more difficult, as is shown more clearly in figure \ref{fig:minTripleProd}. Though other reactions that have been seriously considered (e.g. D-Li-6, He-3-He-3, p-Li-6, p-Be-9, p-B-11) have reasonably low point triple products compared to the rest. This gives confidence that it is a reasonable metric for comparison. Additionally, we see that the ion energies at the minimum point triple product have the right relative values to one another. However, they are systematically higher than the {\it temperatures} that detailed analyses indicate is optimal. This is understandable because a Maxwellian velocity distribution, unlike the delta function we assumed, has a long tail at high energy that has an outsized contribution to fusion energy production. Moreover, a Maxwellian distribution with an average particle energy of $E_{i}$ has a temperature of $T_{i} = 2 E_{i} / 3$. As a consequence, the fusion to bremsstrahlung power density ratios are lower than more realistic treatments \cite{RiderAltFuels1995}. However, their values relative to one another are still illuminating.

With the help of figures \ref{fig:BEcurve} and \ref{fig:possibleReactions}, we can now assess if it is possible to beat the specific energy of the catalyzed D-D+D fuel cycle. To do so, we have two options\footnote{We will not consider muon-catalyzed fusion because creating muons requires a lot of energy and they live for just 2.2 microseconds. We will also ignore the possibility of three-body fusion (e.g. the triple-alpha process) due to its improbability.}:
\begin{enumerate}[label=(\arabic*)]
	\item use inputs with an average mass per nucleon that is greater than hydrogen-2 and/or
	\item make products with an average mass per nucleon that is less than helium-4.
\end{enumerate}
Additionally, we can divide the methods to accomplish either of these into two strategies:
\begin{enumerate}[label=(\Alph*)]
	\item directly fuse input nuclei in a chain to produce larger nuclei or
	\item use nuclei as catalysts to enable a catalytic fusion cycle.
\end{enumerate}
We will consider each of the four combinations in turn.

First, we will look at option 1A: directly fusing inputs with an average mass per nucleon that is greater than hydrogen-2. Consulting figure \ref{fig:BEcurve}, we see that the only isotope that can enable this is hydrogen-1. The free neutron is not naturally occurring as it has a half-life of just 10 minutes, so it cannot be used as an input. Stars directly fuse two hydrogen-1 nuclei into hydrogen-2 as part of the proton-proton chain, but this reaction has a cross-section so small that it has never been experimentally measured. More broadly, figure \ref{fig:possibleReactions} shows that it not feasible to fuse hydrogen-1 with anything smaller than lithium-6. This appears to rule out option 1A because the combination of hydrogen-1 and lithium-6 has an average mass per nucleon of 934.5 MeV/c\textsuperscript{2}, considerably smaller than that of hydrogen-2. Strictly speaking, it could be possible to use a single large nucleus in the first reaction and then repeatedly fuse the products with enough small nuclei for the average mass per nucleon of the reactants become large enough. However, this does not look possible. Even if you could do the reaction $10 \elem{H}{1} + \elem{Li}{6} \rightarrow \elem{O}{16}$, you still wouldn't exceed the specific energy of the catalyzed D-D+D cycle. Such a reaction is not possible because p-F-19 is the only feasible reaction that produces an nucleus with a mass per nucleon below oxygen-16 and it is not possible to produce fluorine-19 with any feasible reaction.

Next, we will consider option 2A: directly fusing inputs to make products with an average mass per nucleon that is less than helium-4. Figure \ref{fig:BEcurve} shows that, because of the exceptional stability of helium-4, it is not enough to just create a larger nucleus --- we must create a much larger nucleus. Specifically, carbon-12 is the smallest nucleus that has a lower mass per nucleon than helium-4. Producing the nuclei in-between (instead of helium-4) would actually decrease specific energy. The difficulty posed by this is exacerbated by the lack of a nucleus with a mass number of 5. The longest-lived nucleus with $A=5$ is helium-5, which has a half-life shorter than $10^{-21}$ seconds \cite{BaumNuclideChart2002}. Because of this gap, it is impossible to go past helium-4 by directly fusing inputs. In stars, this gap is bridged by the triple-alpha process, which fuses three helium-4 nuclei into a single carbon-12 nucleus. However, three-body fusion is far too rare to accomplish on human scales. Looking at figure \ref{fig:possibleReactions}, we see no alternatives. No reaction in the first four rows of figure \ref{fig:possibleReactions} produces a nucleus that is larger than helium-4. Moreover, the easiest reaction in the fifth row is T-He-4, which is a billion times more difficult than D-T. There are reactions that use nuclei with $A > 5$ that do produce carbon-12 or larger nuclei (e.g. reacting D-B-11). However, using these heavier nuclei as input is unworkable because it decreases the average mass per nucleon of the reactants by too much (as discussed in option 1A). Thus, we cannot use option 2A.

Now we will consider the possibility of a catalytic fusion cycle. In a catalytic cycle, large nuclei are used as inputs to fusion reactions, but are later regenerated by subsequent reactions to eliminate them as inputs to the fuel cycle. Such catalytic cycles play an important role in stellar fusion and some have been proposed for terrestrial fusion \cite{McNallyChainReactionsI1971}. It is important to note that using a catalyst with $A<5$ does not help us accomplish our goal due to the discussions of options 1A and 2A. Additionally, no reaction in the entirety of figure \ref{fig:possibleReactions} produces more than one nuclei with $A>5$. {\it This fact directly means that option 2B is impossible.} You cannot use catalysts to build products larger than helium-4, because you'll always need the largest product to regenerate the catalyst. This also means that any reaction with an $A>5$ reactant, but without an $A>5$ product cannot be used in a catalytic cycle. It becomes impossible to regenerate the heavy reactant. This is because (due to the discussion of option 2A), you can only bridge the $A=5$ gap with the help of a larger nucleus and you've just lost it. Similarly, you cannot use any reaction where both reactants have $A>5$. It is not possible to regenerate both and you cannot tolerate having one as an input because of their low mass per nucleon.

With this context, we consider our last option, 1B: using a catalyst to construct helium-4 with hydrogen-1 as an input. This is what is accomplished by the CNO cycle that occurs in stars. However, it relies on fusing protons with carbon-12, carbon-13, nitrogen-14, and carbon-15. All but one of these reactions are far beyond our capabilities. However, we should still search for an easier way. We know that we must start by fusing hydrogen-1 with something that has a product with $A>5$ (in order to have hope of closing the cycle). We see six options that do this and are at least remotely feasible: beryllium-9, boron-10, nitrogen-15, oxygen-18, fluorine-18, and fluorine-19. Let's examine them in turn.

The beryllium-9 reaction does produce a nucleus with $A>5$, but only in one of two branches. Moreover, reconstructing the beryllium-9 requires that only the neutron-producing branch of the D-Li-6 reaction be used (and you must wait 50 days for beryllium-7 to decay into lithium-7). However, if this can be done, it enables the overall reaction $\elem{H}{1} + \elem{H}{2} + \elem{H}{3} \rightarrow \elem{He}{4} + 2 \elem{n}{1}$. The hardest constituent reaction of this cycle has a point triple product of $1.6 \times 10^{25}$ keV-s/m\textsuperscript{3}. Of course, this is almost a hundred times harder than D-D, but we're considering it to be in the realm of possibility. Another difficulty is obtaining the tritium without ruining the specific energy. The only possibility is to add the branch of D-D fusion that produces tritium, enabling tritium to be eliminated as an input. Then, deuterium breeding using normal hydrogen (i.e. equation \refEq{eq:Dbreeding}) can be used to eliminate one of the input deuterons and we can achieve the reaction $2 \elem{H}{1} + \elem{H}{2} \rightarrow \elem{He}{4}$. This fuel cycle has a specific energy of 6.3 MeV/AMU, somewhat higher than catalyzed D-D+D! Unfortunately, apart from all the other difficulties, we have assumed that we can exclude specific branches of three different reactions at will. Simply put, this does not appear possible to the degree that we require. The cross-sections for the different branches have similar magnitudes and depend on temperature in the same way.

The boron-10 reaction can also enable the $\elem{H}{1} + \elem{H}{2} + \elem{H}{3} \rightarrow \elem{He}{4} + 2 \elem{n}{1}$ cycle. Accomplishing this requires the T-Li-7 and D-Be-9 reactions (as well as waiting for beryllium-7 to decay into lithium-7). While this involves only one reaction with branches that must be excluded, it starts from the very hard p-B-10 reaction. Regardless, attaining the $2 \elem{H}{1} + \elem{H}{2} \rightarrow \elem{He}{4}$ fuel cycle still requires tritium, so the neutron-producing branch of the D-D reaction must be blocked. Thus, starting with boron-10 does not appear any more feasible than starting with beryllium-9.

The last four options are fairly similar. Fusing hydrogen-1 with nitrogen-15, which is part of the solar CNO cycle, produces carbon-12. Unfortunately, carbon-12 is so stable that it doesn't react with anything. Oxygen-18 produces nitrogen-15, which can only be split further, so the cycle cannot be closed. Fluorine-18 produces oxygen-15, which decays into nitrogen-15, but neither can be used to produce a larger nuclei. Finally, fluorine-19 produces oxygen-16 or neon-20, both of which are so stable that they don't react with anything.

In conclusion, we've taken quite a granular look at the possible fusion reactions and been unable to find a viable fuel cycle with a specific energy larger than that of catalyzed D-D+D. Moreover, we've discussed some arguments that suggest it is impossible, given the known fusion cross-sections. The primary features can be summarized as:
\begin{itemize}
	\item There are no nuclei with $A = 5$.
	\item Hydrogen-1 cannot be fused with any other nucleus with $A < 5$.
	\item There is no viable reaction between two nuclei with $A < 5$ that creates a nuclei with $A > 5$,
	\item Using a nucleus with $A > 5$ as an input decreases the mass per nucleons of the reactants by too much to allow a high specific energy.
	\item A nucleus with $A > 5$ cannot be used as a catalyst because there are few viable fusion reactions involving large nuclei, so it cannot be regenerated without the ability to block certain branches of several reactions.
\end{itemize}

\section{Space propulsion}
\label{sec:propulsion}

Lastly, a particularly important application of a catalyzed D-D+D fuel cycle could be space propulsion. In space travel, two of the most important performances metrics are specific {\it energy} and specific {\it momentum}. Here specific momentum is defined to be the total directed momentum of the exhausted particles divided by the mass of the input particles, so it could also be termed an effective exhaust velocity. Specific {\it energy} is more important for propulsion schemes with dedicated propellant (e.g. \cite{EnglertFirstDHe3spacecraft1962, HiltonFusionPropulsion1964, MileyDHe3propulsion2007}), while specific {\it momentum} is key for schemes that directly exhaust the fusion products as propellant (e.g. \cite{HiltonFusionPropulsion1964, SantariusDHe3Spacecraft1992}). Having dedicated propellant increases the total amount of thrust, but also increases the total combined mass of fuel/propellant that is required to achieve a given spacecraft velocity. For example, the rocket equation shows that a spacecraft must be nearly 90\% fuel/propellant to reach a speed of double its specific momentum. On the other hand, high thrust is desirable for short journeys and to navigate the gravitational fields within a star system.

Many studies \cite{EnglertFirstDHe3spacecraft1962, HiltonFusionPropulsion1964, SantariusDHe3Spacecraft1992} have concluded that a pure D-He-3 fuel cycle is preferable to both catalyzed D-D and D-T. This is sensible because, without deuterium breeding, the D-He-3 reaction is optimal for both specific energy (see table \ref{tab:specEnergy}) and specific momentum. However, we have just learned that deuterium breeding enables the specific energy of catalyzed D-D to surpass D-He-3. Thus, it has the potential to be useful. In this section, we will first calculate the impact of deuterium breeding on specific momentum and then discuss the overall viability of catalyzed D-D+D propulsion.

Without analysis, it is not clear if a catalyzed D-D+D fuel cycle could surpass the specific momentum of D-He-3. A D-He-3 fusion device would release virtually all of its energy in the form of charged particles, which can be efficiently redirected into a beam to maximize thrust. This enables D-He-3 fuel to achieve a specific momentum approaching 7.1\% of the speed of light. On the other hand, the catalyzed D-D+D fuel cycle is theoretically capable of a specific momentum of 11.3\% of the speed of light. However, this assumes that the 23.8 MeV released by the chain of reactions underlying equation \refEq{eq:catalyzedDDwithBreeding} can be entirely transferred to the helium-4 product. In practice, this does not appear possible because much of this energy is given to neutrons, which is difficult to use efficiently. Nevertheless, how close we can get involves a complex design optimization that depends strongly on the practicality and efficiency of various technological components. Given that net energy production with any fusion fuel has yet to be achieved, a detailed investigation is premature. Instead, we will present a few idealized examples to convey a sense of the possibilities and limitations.

\begin{figure}
	\centering
	\includegraphics[width=\textwidth]{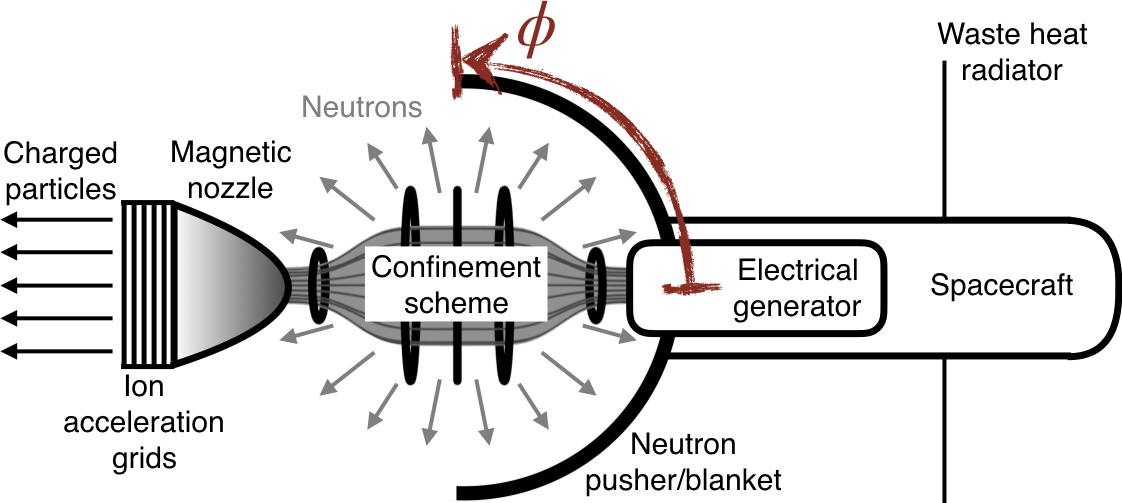}
	\caption{A cartoon fusion-powered spacecraft with neutron pusher/blanket extending over $\phi \in [0, \pi/2]$ and the full $[0, 2 \pi]$ in azimuthal angle. As an example, a magnetic mirror is used as the confinement scheme.}
	\label{fig:propulsionGeo}
\end{figure}

In conventional catalyzed D-D fusion the fuel cycle can be written as
\begin{align}
  6 ~ \elem{H}{2} \to& \elem{He}{4} ~ \text{(3.5 MeV)} + \elem{H}{1} ~ \text{(3.02 MeV)} + \elem{n}{1} ~ \text{(14.1 MeV)} \\
  & + \elem{He}{4} ~ \text{(3.6 MeV)} + \elem{H}{1} ~ \text{(14.7 MeV)} + \elem{n}{1} ~ \text{(2.45 MeV)} + \text{1.83 MeV} , \nonumber
\end{align}
where the kinetic energy carried by each particle is indicated in the parentheses and the energy that is carried by intermediate products (which remains on-board) is indicated by the last term. This shows that 38\% of the total energy is carried off by neutrons and 4\% is carried by the \elem{H}{3} and \elem{He}{3} fusion products, which must remain in the spacecraft to fuse. While we assume that charged products can be exhausted optimally, we will assume that neutrons are born isotropically and cannot be redirected. Consequently, only $1/4$ of the neutron's total momentum can be utilized for thrust. This is accomplished by extending a barrier from $0$ to $\pi/2$ in the zenith angle $\phi$ (see figure \ref{fig:propulsionGeo}) in order to block all of the neutrons traveling forward and allow all of the neutrons traveling backwards to freely stream away. Even with such geometry, the normal catalyzed D-D fuel cycle achieves a specific momentum of only 5.6\% of the speed of light.

Introducing a blanket of normal hydrogen that extends over the full $\phi \in [0, \pi]$ enables deuterium breeding and achieves the fuel cycle
\begin{align}
  4 ~ \elem{H}{2} \to& \elem{He}{4} ~ \text{(3.5 MeV)} + \elem{He}{4} ~ \text{(3.6 MeV)} + \text{40.5 MeV} , \label{eq:fullDbreeding}
\end{align} 
which is equivalent to equation \refEq{eq:catalyzedDDwithBreeding}. Even assuming that the helium-4 is exhausted optimally, this fuel cycle actually has a lower specific momentum than normal catalyzed D-D fusion --- only 4.4\% of the speed of light. This is because the normal hydrogen produced by fusion, which previously carried significant amounts of thrust, is now being kept on-board to breed deuterium. However, we see that there is a tremendous amount of energy now available on the spacecraft. This is bad if waste heat removal is difficult \cite{SantariusDHe3Spacecraft1992, MaslenEarlyPropulsion1959} or good if additional power is useful to run the confinement scheme/other on-board systems. For example, one could imagine using this power to directly augment the thrust by adding an ion acceleration stage at the end of the fusion thruster. If the on-board energy could be transferred to the outgoing helium-4 with an overall efficiency of just 30\%, a specific momentum of 7.2\% of the speed of light could be achieved. This would require accelerating particles from 3.5 MeV to 9.5 MeV, but would enable the catalyzed D-D+D fuel cycle to surpass the specific momentum of D-He-3. An overall combined fusion product-to-acceleration efficiency of 30\% does not appear impossible. Much of the on-board energy is carried by charged particles, which can theoretically be directly converted to electricity at high efficiency, while the rest could go through a steam cycle. The feasibility and efficiency of high-energy ion acceleration is very uncertain, but low-energy ion thrusters currently have efficiencies as high as 80\% \cite{ChoueiriIonThrustEff2009}. Note that, in the limit of a 100\% overall fusion product-to-acceleration efficiency, the theoretical maximum of 11.3\% of the speed of light is achieved.

Without ion acceleration, it is better to exhaust the normal hydrogen produced by fusion and breed deuterium with hydrogen that has been brought along. This gives the fuel cycle
\begin{align}
  4 ~ \elem{H}{2} + 2 \elem{H}{1} \to& \elem{He}{4} ~ \text{(3.5 MeV)} + \elem{H}{1} ~ \text{(3.02 MeV)} \label{eq:fullDbreedingCarryingH} \\
  &+ \elem{He}{4} ~ \text{(3.6 MeV)} + \elem{H}{1} ~ \text{(14.7 MeV)} + \text{22.8 MeV} , \nonumber
\end{align}
which has a specific momentum of 6.1\% of the speed of light --- better than the standard catalyzed D-D, but less than D-He-3. 
Alternatively, for missions involving very long travel times, one could modify the catalyzed D-D cycle somewhat by waiting for the tritium to decay into helium-3 before fusing it \cite{KesnerHeliumCatDD2003}. When combined with deuterium breeding it could enable the fuel cycle
\begin{align}
5 ~ \elem{H}{2} \to& 2 ~ \elem{He}{4} ~ \text{(3.6 MeV)} + 2 ~ \elem{H}{1} ~ \text{(14.7 MeV)} + \text{9.52 MeV} .
\end{align}
This is very similar to the standard D-He-3 reaction and has the same specific momentum of 7.1\% of the speed of light. However, it has the advantages of only requiring deuterium as input and generating extra on-board energy. The disadvantages include a massive breeding blanket/radiator, waiting ${\sim} 12$ years for tritium to decay, and an overall increased level of technical complexity.

Though even more speculative, what would significantly improve the specific momentum would be to collect normal hydrogen along the flight path. This would allow a fuel cycle to directly exhaust the hydrogen-1 produced by fusion, without having to bring it along as an input. For example, eliminating the input hydrogen-1 in equation \refEq{eq:fullDbreedingCarryingH} boosts the specific momentum from 6.1\% to 7.6\% --- somewhat larger than D-He-3. Moreover, there is still 22.8 MeV of on-board energy that could be used to power the hydrogen collection scheme and/or for ion acceleration. Using ion acceleration with a fusion product-to acceleration efficiency of 30\% could boost the specific momentum to 8.9\% of the speed of light.

The idea of collecting hydrogen is similar to the Bussard ramjet \cite{BussardRamjet1960, WhitmireRamjet1975}, which uses large electromagnetic fields to gather particles from kilometers around\footnote{Catalyzed D-D+D fusion could also improve the performance of a related concept: the RAIR \cite{BondRAIR1974, PowellRAIR1975, JacksonRAIR1980}. On-board fuel is used to power the spacecraft, but the propellant is gathered from interstellar space, so the specific energy (rather than momentum) is what matters.}. While the design detailed in this work requires a similar number of particles to be collected, it is much more practical than the Bussard ramjet because it does not depend on p-p fusion (which has an extremely small cross-section \cite{AdelbergerSolarCrossSections1998}), nor the collection of deuterium (which is tens of thousands of times less abundant than normal hydrogen \cite{OmearaDeuteriumAbundance2001}). The catch is that getting the collected hydrogen into the breeding blanket would require it to be accelerated to the speed of the spacecraft. During the spacecraft's acceleration, any savings that were gained by not bringing the fuel would be entirely canceled by drag against the particles that must be gathered. However, during deceleration it could be doubly helpful --- the drag would further decelerate the spacecraft, while the collected hydrogen could be used to breed deuterium. 


In summary, we see that adding deuterium breeding can substantially improve the specific momentum of a catalyzed D-D propulsion system, potentially enabling it to surpass D-He-3. 
Importantly, the primary advantages of a catalyzed D-D fuel cycle are retained. Catalyzed D-D uses fuel that is extremely abundant and readily available on Earth and produces plasmas that have a higher power density and ignite at a lower temperature than D-He-3 \cite{MillsCatDD1971, MileyDDtokamaks1977}. However, we have seen that a D-D propulsion system and the techniques needed to improve its performance carry the drawbacks of increased technical complexity and the need for massive components (e.g. neutron shielding, waste heat radiators, ion acceleration grids, hydrogen collection systems). Thus, the ultimate practicality of a fuel cycle will depend strongly on factors like how light-weight these components are and how well they perform.

\section{Conclusions}
\label{sec:conclusions}

This work has tried to answer the question ``What fuel source can provide the most energy with the least mass?'' In doing so, we have considered an addition to the catalyzed D-D fusion fuel cycle. By surrounding the reactor with a blanket containing normal hydrogen, the neutrons from fusion can be used to breed deuterium and eliminate a substantial fraction of the fuel needed to run the power plant. Given the ability to achieve a catalyzed D-D cycle, this addition appears straightforward to accomplish --- one could simply immerse the reactor in a tank of liquid water. By studying experimentally measured fusion cross-sections, we have argued that such a scheme enables the maximum specific energy of any known fuel source. Apart from satisfying our own curiosity, such an answer reveals how to most efficiently use the dominant energy stockpile we have here on Earth and in the universe. Moreover, such a scheme could be useful for any application that requires fuel to be transported. We have analyzed one such application, space propulsion, and seen that a catalyzed D-D thruster using deuterium breeding is theoretically capable of surpassing the performance of D-He-3.

\section*{Acknowledgments}

The author is indebted to J. Parisi and H. de Oliveira for valuable discussions and comments.

\section*{References}
\bibliographystyle{unsrt}
\bibliography{references.bib}

\end{document}